\documentclass{ws-procs9x6}

\begin{document}

\title{SU(3)-breaking effects in hyperon semileptonic decays\\
from lattice QCD}

\author{Silvano SIMULA}

\address{INFN, Sezione Roma 3, Via della Vasca Navale 84, I-00146 Roma, Italy\\
{\it E-mail: simula@roma3.infn.it}}

\begin{abstract}
We present the first quenched lattice QCD study of all the vector and axial form factors 
relevant for the hyperon semileptonic decay $\Sigma^- \rightarrow n~\ell~\nu$.
\end{abstract}


\bodymatter


\vspace{0.5cm}

\indent Recently it has been shown \cite{Kl3} that SU(3)-breaking corrections to the 
$K \to \pi$ vector form factor (f.f.) can be determined from lattice simulations with great 
precision, allowing to reach the percent level of accuracy in the extraction of $V_{us}$ 
from $K_{\ell3}$ decays. An independent way to extract $V_{us}$ is provided by hyperon 
semileptonic decays, and Ref.~\cite{CSW} has shown that it is possible to extract the 
product $|V_{us} \cdot f_1(0)|$ at the percent level from experiments, where $f_1(0)$ 
is the vector f.f.~at zero-momentum transfer.

\indent In Ref.~\cite{hyperon} we have performed a lattice QCD study of SU(3)-breaking 
corrections in all the vector and axial f.f.'s relevant for the $\Sigma^- \rightarrow 
n~\ell~\nu$ decay. Though the simulation has been carried out in the quenched approximation, 
our results represent the first attempt to evaluate hyperon f.f.'s using a non-perturbative 
method based only on QCD.

\indent For each f.f.~we have studied its momentum and mass dependencies, obtaining its value 
extrapolated at $q^2 = 0$ and at the physical point. Our results are collected in Table~\ref{tab:final}.

\begin{table}
\tbl{Lattice results~\cite{hyperon} for the vector and axial form factors at $q^2 = 0$ for the $\Sigma 
\rightarrow n$ transition. The errors do not include the quenching effect.
\vspace{0.1cm}}
{\begin{tabular}{||c||c||}
\hline 
 $f_1(0)$          & $-0.988 \pm 0.029_{\rm lattice} \pm 0.040_{\rm HBChPT}$\\ \hline
 $g_1(0) / f_1(0)$ & $-0.287 \pm 0.052$\\ \hline
 $f_2(0) / f_1(0)$ & $-1.52 \pm 0.81$\\ \hline
 $f_3(0) / f_1(0)$ & $-0.42 \pm 0.22$\\ \hline
 $g_2(0) / f_1(0)$ & $+0.63 \pm 0.26$\\ \hline
 $g_3(0) / f_1(0)$ & $+6.1 \pm 3.3$\\ \hline
\hline
\end{tabular}}
\label{tab:final}
\end{table}

\indent It can be seen that the SU(3)-breaking corrections to $f_1(0)$ have been determined with great 
statistical accuracy in the regime of the simulated quark masses, which correspond to pion masses above 
$0.7$ GeV. The magnitude of the errors reported in Table~\ref{tab:final} is mainly due to the chiral 
extrapolation and to the poor convergence of the Heavy Baryon Chiral Perturbation Theory \cite{Villa}. 
Though within large errors the central value of $f_1(0)$ arises from a partial cancellation between the 
contributions of local terms, evaluated on the lattice, and chiral loops. This may indicate that 
SU(3)-breaking corrections on $f_1(0)$ are moderate, giving support to the analysis of Ref.~\cite{CSW}.

\indent The ratio $g_1(0) / f_1(0)$ is found to be negative and consistent with the value adopted in 
Ref.~\cite{CSW}. In the limit of exact SU(3) symmetry we obtain $[g_1(0) / f_1(0)]_{SU(3)} = -0.269 \pm 
0.047$. This means that SU(3)-breaking corrections are moderate also on this ratio, though it is not 
protected by the Ademollo-Gatto theorem against fist-order corrections.

\indent The weak electricity form factor $g_2$ is found to be non-vanishing because of SU(3)-breaking 
corrections. Our result for $g_1(0) / f_1(0)$ combined with that of $g_2(0) / f_1(0)$ are nicely consistent 
with the experimental result from Ref.~\cite{Sigma}. Our findings favor the scenario in which $g_2(0) / 
f_1(0)$ is large and positive with a corresponding reduced value for $|g_1(0) / f_1(0)|$ with respect to 
the conventional assumption $g_2(q^2) = 0$ based on exact SU(3) symmetry.

\indent Future lattice QCD studies of the hyperon semileptonic transitions should remove the quenched 
approximation and lower the quark masses as much as possible in order to reduce the impact of the 
chiral extrapolation. The accuracy of the ratios $f_2(0) / f_1(0)$, $f_3(0) / f_1(0)$, $g_2(0) / 
f_1(0)$ and $g_3(0) / f_1(0)$ may be improved by implementing twisted boundary conditions for the 
quark fields (see Ref.~\cite{theta_3pt}), which allow to reach values of the momentum transfer 
closer to $q^2 = 0$. Finally, the use of smeared source and sink for the interpolating fields as 
well as the use of several, independent interpolating fields may help in increasing the overlap 
with the ground-state signal, particularly at low values of the quark masses.

\bibliographystyle{ws-procs9x6}

\bibliography{ws-pro-sample}

\begin{thebibliography}{99}

\bibitem{Kl3}
  D.~Becirevic {\it et al.},
  {\em Nucl.\ Phys.\ B} {\bf 705}, 339 (2005);
  {\em Eur.\ Phys.\ J.\ A} {\bf 24S1}, 69 (2005).
\bibitem{CSW}
  N.~Cabibbo {\it et al.},
  {\em Ann.\ Rev.\ Nucl.\ Part.\ Sci.\ } {\bf 53}, 39 (2003).
\bibitem{hyperon}
  D.~Guadagnoli {\it et al.},
  hep-ph/0606181;
  {\em Nucl.\ Phys.\ Proc.\ Suppl.\ }  {\bf 140}, 390 (2005);
  {\em PoS LAT2005}, 358 (2005).
\bibitem{Villa} 
  G.~Villadoro,
  {\em Phys.\ Rev.\ D} {\bf 74}, 014018 (2006).
\bibitem{Sigma}
  S.~Y.~Hsueh {\it et al.},
  {\em Phys.\ Rev.\ D} {\bf 38}, 2056 (1988).
\bibitem{theta_3pt}
  D.~Guadagnoli {\it et al.},
  {\em Phys.\ Rev.\ D} {\bf 73}, 114504 (2006).

\end{thebibliography}

\end{document}